\documentclass[]{iopart}

\expandafter\let\csname equation*\endcsname\relax
\expandafter\let\csname endequation*\endcsname\relax
\usepackage{graphicx}
\usepackage{epstopdf}
\usepackage{cite}
\usepackage[perpage,marginal]{footmisc}
\usepackage{etoolbox}
\usepackage{marvosym}
\usepackage{upgreek}
\usepackage{pdflscape}
\usepackage[unicode]{hyperref}
\usepackage[all]{hypcap} 
\usepackage{amsmath,amsfonts,amssymb}
\usepackage{color}
\usepackage[usenames,dvipsnames]{xcolor}
\usepackage{bm}
\usepackage{longtable}
\usepackage[T1]{fontenc}
\usepackage[utf8]{inputenc}
\usepackage{verbatim}
\usepackage{pifont}
\usepackage{yfonts}
\usepackage{url}
\usepackage{multirow}
\usepackage[normalem]{ulem}
\makeatletter
\def\url@leostyle{
  \@ifundefined{selectfont}{\def\UrlFont{\sf}}{\def\UrlFont{\small\ttfamily}}}
\makeatother
\usepackage[normalem]{ulem}
\usepackage{fancyhdr}
\makeatletter
\renewcommand\footnoterule{
  \kern-3\p@
  \hrule\@width2.5cm
  \kern2.6\p@}
\makeatother
\numberwithin{equation}{section}
\numberwithin{figure}{section}
\definecolor{linkcolor}{rgb}{0.0,0.3,0.5}
\definecolor{urlcolor}{rgb}{0.27,0.55,0.}
\definecolor{funcolor}{rgb}{0.65, 0.16, 0.16}
\hypersetup{colorlinks=true,linkcolor=linkcolor,citecolor=linkcolor,filecolor=linkcolor,urlcolor=linkcolor}

\begin{document}

\pagestyle{fancy}
\lhead{}
\chead{}
\rhead{D. Gerosa \emph{et al.}}
\lfoot{}
\cfoot{\thepage}
\rfoot{}

\begin{center}
\title{On the equal-mass limit of precessing black-hole binaries}
\end{center}

\author{
Davide~Gerosa$^{*\,1,2}$,
Ulrich~Sperhake$^{2,1,3}$
and Jakub Vo\v{s}mera$^{4,2}$
}
\address{$^{1}$~TAPIR 350-17, California Institute of Technology, 1200 E California Boulevard, Pasadena, CA 91125, USA}
\address{$^{2}$~Department of Applied Mathematics and Theoretical Physics, Centre for Mathematical Sciences, University of Cambridge, Wilberforce Road, Cambridge CB3 0WA, UK}
\address{$^{3}$~Department of Physics and Astronomy,
The University of Mississippi, University, MS 38677, USA}
\address{$^{4}$~Institute of Physics AS CR, Na Slovance 2, Prague 8, Czech Republic}
\address{*~Einstein Fellow}

\ead{\href{mailto:dgerosa@caltech.edu}{dgerosa@caltech.edu}}

\begin{abstract}
We analyze the inspiral dynamics of equal-mass precessing black-hole
binaries using  multi-timescale techniques. The orbit-averaged
post-Newtonian evolutionary equations admit two constants of motion
in the equal-mass limit, namely the magnitude of the total spin $S$
and the effective spin $\xi$. This feature makes the entire dynamics
qualitatively different compared to the generic unequal-mass case,
where only $\xi$ is constant while the variable $S$ parametrizes the
precession dynamics. For fixed individual masses and spin magnitudes,
an equal-mass black-hole inspiral is
uniquely characterized by the two parameters
$(S,\xi)$: these two numbers completely determine the
entire evolution under the effect of radiation reaction. In particular,
for equal-mass binaries we find that  (i) the black-hole binary spin
morphology is constant throughout the inspiral, and that (ii) the
precessional motion of  the two black-hole spins about the total
spin takes place on a longer timescale than the precession of the
total spin and the orbital plane about the  total angular momentum.
\end{abstract}
\mathindent = 0pc

\section{Introduction}

In the framework of general relativity,
the dynamics of black-hole (BH) binaries and their emitted
gravitational-wave (GW) signals are determined by the masses and spins  of
the inspiralling BHs.
The binary's total mass $M=m_1+m_2$ primarily sets the GW  frequency,
and therefore determines the required detection technique. Ground-based
interferometers are sensitive to BH binaries with masses
$\mathcal{O}(1-100) M_\odot$
\cite{2015CQGra..32g4001L,2015CQGra..32b4001A}, space GW missions
will be most sensitive to BHs of $M\sim \mathcal{O}(10^4-10^7)
M_\odot$ \cite{2013arXiv1305.5720C}, while Pulsar Timing Arrays
target the detection of GWs from even more massive binaries of
$\mathcal{O}(10^8-10^{10}) M_\odot$
\cite{2013PASA...30...17M,2013CQGra..30v4009K,2009arXiv0909.1058J}. The
mass ratio $q=m_2/m_1\leq 1$ directly enters the BH dynamics and
the quantity most accurately determined in observations of BH
inspirals is the binary's chirp mass $M_c= M[q/(1+q)]^{3/5}$
which sets the phase of the emitted GWs
\cite{1995PhRvD..52..848P,2009LRR....12....2S}.  The BH spins
$\mathbf{S_1}$ and $\mathbf{S_2}$, although more difficult to
measure, also directly affect the GW signal. Specifically, the spin
components aligned with the orbital angular momentum affect the
coalescence time as more (if spins are aligned) or less (if
spins are anti-aligned) angular momentum is shed before merger
\cite{2006PhRvD..74d1501C}. This effect may be viewed as part of
general relativity's tendency to cloak spacetime singularities
inside horizons according to Penrose's cosmic censorship conjecture
\cite{1969NCimR...1..252P,1997gr.qc....10068W} as excessive angular momentum of
the post-merger BH would imply a naked singularity.  In the presence
of non-vanishing spin components perpendicular to the orbital angular
momentum,
precession due to relativistic spin-spin and spin-orbit coupling
introduces characteristic modulations in the emitted chirp
\cite{1995PhRvD..52..821K,1994PhRvD..49.6274A}. Future observation
of these patterns may help in
determining the formation channel of binary BHs
\cite{2013PhRvD..87j4028G,2016ApJ...832L...2R,2016ApJ...818L..22A}. The
BH spins also play a crucial role in determining the final properties
of the BH remnant
\cite{2009CQGra..26i4023R,2010PhRvD..81h4054K,2016ApJ...825L..19H},
especially its recoil velocity
\cite{2007PhRvL..98w1101G,2007PhRvL..98w1102C,2010ApJ...715.1006K}.
In terms of gravitational-wave source modelling, BH spins increase the
number of source parameters by six, significantly increasing
the complexity of the systems; see, for instance,
\cite{2016arXiv161103703B,2016PhRvD..93d4007K,2014PhRvD..90d4018D,2016arXiv160603117C}, also for attempts
to simplify the task. It is highly desirable, in this context, to
dissect, in so far as possible using analytic means, the complicated
morphology of spin precession and classify its key features.

Spin precession influences the binary dynamics on timescales $t_{\rm
pre}\propto r^{5/2}$ (where $r$ is the binary separation)
\cite{1994PhRvD..49.6274A,1995PhRvD..52..821K}. In the post-Newtonian
(PN) regime $r/M\gg 1$ and $t_{\rm pre}$ is (i)
much longer than the orbital period $t_{\rm orb} \propto r^{3/2}$,
and (ii) much shorter than the inspiral timescale $t_{\rm
RR}\propto r^{4}$ \cite{1964PhRv..136.1224P}. The resulting hierarchy
\begin{align}
  t_{\rm orb}\ll t_{\rm pre} \ll t_{\rm RR}
\end{align}
turns out to be a very powerful tool to study the binary dynamics:
different processes (namely orbital motion, precession and inspiral) can be
modelled on their respective timescales by averaging over quantities
varying on shorter times and keeping constant those variables that only
evolve over the longer time scales. The resulting equations can then
be reassembled as a complete formalism
using a
quasi-adiabatic approach. This idea is at the heart of the decades-old
orbit-averaged  formulation of the BH binary dynamics
\cite{1964PhRv..136.1224P,1994PhRvD..49.6274A,1995PhRvD..52..821K}, as
well as the new precession-averaged PN approach
\cite{2015PhRvL.114h1103K,2015PhRvD..92f4016G}.

In this paper, we complement the analysis of
\cite{2015PhRvL.114h1103K,2015PhRvD..92f4016G} by studying in detail
equal-mass systems ($q=1$). At first glance, this may appear as a
predominantly academic exercise, but it is also of practical importance
for at least four reasons.

\begin{enumerate}
\item
As we will discuss at greater length below,
the subset of $q=1$ binaries behaves
qualitatively different in several regards relative
to the generic unequal-mass case
studied in \cite{2015PhRvL.114h1103K,2015PhRvD..92f4016G} because
of the existence of an additional constant of motion
(Sec.~\ref{constantsofmotion}). This phenomenon also manifests itself
at a formal level: merely setting $q=1$ in the mathematical
framework developed in \cite{2015PhRvL.114h1103K,2015PhRvD..92f4016G}
leads to singular expressions in various places and, hence,
does not directly predict the dynamics of equal-mass binaries.
\item Even though the behaviour of $q=1$ binaries is ultimately
derived from the formalism developed in
\cite{2015PhRvL.114h1103K,2015PhRvD..92f4016G}, the emerging
picture is of such remarkable simplicity that it serves as
an ideal pedagogical introduction and motivation for readers to
venture on to the more complex spin-precession formalism of
the cited work.
\item
During the first years after the numerical relativity breakthroughs
\cite{2005PhRvL..95l1101P,2006PhRvL..96k1102B,2006PhRvL..96k1101C} the majority of
numerical BH studies focussed on equal-mass binaries and, to this day,
equal-mass binaries have frequently been used as testbeds for BH evolutions
\cite{2007arXiv0710.1338P,2010RvMP...82.3069C,2012CQGra..29l4004P,2015LRR....18....1C,2015CQGra..32l4011S}.
This choice is quite natural for several reasons (for instance,
the additional symmetry allows for reduced computational domains
and the merger dynamics probe the most strongly non-linear regime),
but it involves a small risk that observations made for
this particular class of binaries be mistaken as generically valid.
Our study provides a cautionary statement in this regard, as we
indeed identify characteristic features that hold for equal-mass
systems and {\it only} for equal-mass systems: the constancy of
the total spin magnitude $S$ and a difference in timescale between
the precession of the individual spins and that of the orbital plane.
\item
Vice versa, the extraordinarily simple behaviour of the equal-mass
case may yield valuable insight into characteristic features
of BH binaries that, while not exactly valid for $q\ne 1$, may still
hold approximately and thus contribute to our understanding of general
systems.
For instance, if BH binaries with mass ratio reasonably close to
unity are found, the additional constant of motion stressed in
Sec.~\ref{constantsofmotion} will still be conserved \emph{in
practice} at some approximate level and may thus be useful in the
modeling of GW signals.
We note, in this context, that the
first GW observations are all compatible with equal-mass BH binary sources
within a 90\% credibility interval
\cite{2016PhRvX...6d1014A,2016PhRvL.116x1102A,2016PhRvL.116x1103A,
2016PhRvX...6d1015A}.
\end{enumerate}

The remainder of this work is organized as follows.
Our calculations are carried out in Sec.~\ref{peculiarities}; results are illustrated in Sec.~\ref{results} and conclusions drawn in Sec.~\ref{discussion}.

\section{Peculiarities of the equal-mass case}
\label{peculiarities}
\begin{figure}\centering
  \includegraphics[width=0.55\textwidth]{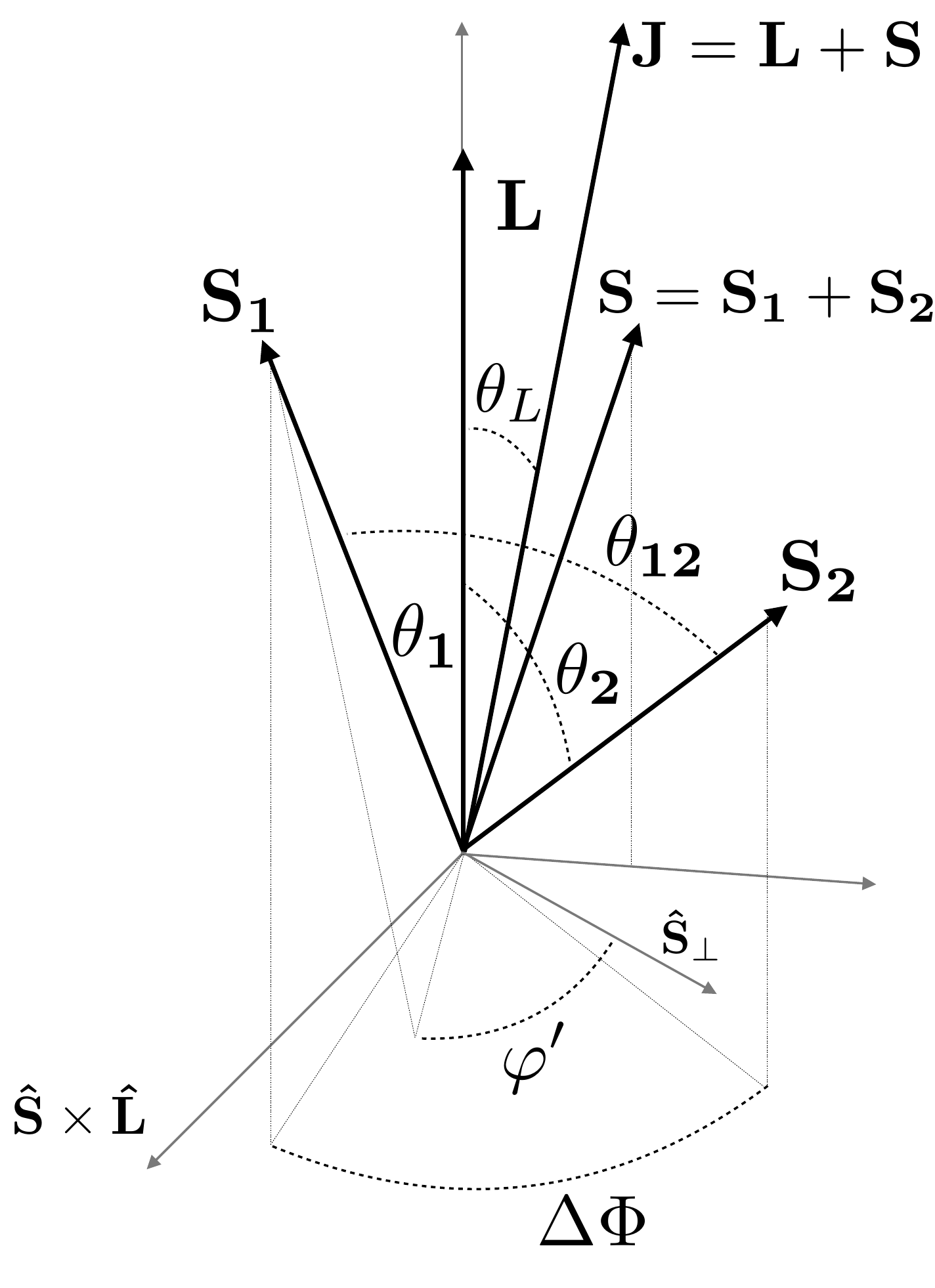}
  \caption{Vectors and angles describing the dynamics of spinning BH binaries. The directions of the two spins $\mathbf{S_1}$ and $\mathbf{S_2}$ with respect to the orbital angular momentum $\mathbf{L}$ are described in terms of the angles $\mathbf{\theta_1}$, $\theta_2$, $\theta_{12}$ and $\Delta\Phi$, cf. Eqs.~(\ref{eq:cth1}-\ref{deltaphidef}). The orientation of the total angular momentum $\mathbf{J}$ relative to $\mathbf{L}$ is specified by the angle $\theta_L$. The angle $\varphi'$ used in Sec.~\ref{orbitalplaneandspinprec} to describe the spin dynamics is measured in the plane orthogonal to $\mathbf{S}$. }
  \label{angles}
\end{figure}

According to the spin-precession formalism developed in
\cite{2015PhRvL.114h1103K,2015PhRvD..92f4016G}, the evolution
of the BH spins is conveniently split into dynamics occurring
on the precession time scale $t_{\rm pre}$ and those happening on the much
longer radiation reaction time scale $t_{\rm RR}$. A key simplification
arises from the fact that the projection of the effective spin along the
orbital angular momentum,
\begin{align}
  \xi= \frac{1}{M} \left( \frac{\mathbf{S_1}}{m_1} +
  \frac{\mathbf{S_2}}{m_2}\right)\cdot \mathbf{\hat L}\,,
\label{defxi}
\end{align}
is a constant of motion of the orbit-averaged 2PN spin-precession
equations and 3.5PN radiation-reaction equation \cite{2001PhRvD..64l4013D,2008PhRvD..78d4021R}; at this order,
$\xi$ is constant on
both time scales $t_{\rm pre}$ and $t_{\rm RR}$.
The binary dynamics on the precession time scale is parametrized by the magnitude $S$ of the total spin $\mathbf{S}=\mathbf{S}_1+\mathbf{S}_2$ while the secular evolution under the effect of radiation reaction is encoded into the total angular momentum $\mathbf{J}=\mathbf{L}+\mathbf{S}$.
A schematic view of these vectors and the angles between them is shown in Fig.~\ref{angles}.

\subsection{Constants of motion}

\label{constantsofmotion}
Constants of motion play a crucial role in this story. Let us 
consider an equal-mass BH binary, with mass ratio $q=m_2/m_1=1$
and total mass $M=m_1+m_2$ on a quasi-circular orbit. The magnitudes of the BH spins
$S_i=m_i^2\chi_i = M^2 \chi_i / 4$ ($i=1,2$) are described in terms of the
dimensionless Kerr parameter $0\leq \chi_i \leq 1$, while
the magnitude of the orbital angular momentum $L$ is given (at the PN order here considered) in terms  of the
binary separation $r$ by the Newtonian relation $L=m_1m_2 \sqrt{r
/M}= \sqrt{r M^3}/4$.
In this case, the effective spin projection $\xi$
(\ref{defxi}) becomes
\begin{equation}
  \xi= 2 \frac{\mathbf{S}\cdot\mathbf{\hat L}}{M^2}
  ~~~~~\Rightarrow~~~~~
  |\xi| \le \frac{2S}{M^2}\,.
\label{defxiq1}
\end{equation}

Both the total angular momentum $\mathbf{J}$
and the magnitude of the orbital angular momentum $L$ are constant
on the precessional timescale $t_{\rm pre}$, because
GWs only dissipate energy and momentum on $t_{\rm RR}$. The magnitude of the total spin $S$, on the other
hand, may vary on $t_{\rm pre}$ under the effect of relativistic spin-spin
and spin-orbit couplings and was used in
\cite{2015PhRvL.114h1103K,2015PhRvD..92f4016G} to parametrize the
precession dynamics for $q\neq 1$ binaries.
It was noted,
however, that there exist a few
special configurations where $S$ is constant, such as
Schnittman's spin-orbit resonances \cite{2004PhRvD..70l4020S} and
stable binary configurations with aligned spins
\cite{2015PhRvL.115n1102G}.

We find here the first qualitative difference in the behaviour
of equal-mass binaries;
for $q=1$, the total spin magnuitude $S$ is {\it always} constant.
This can be seen as follows.
The standard orbit-averaged spin precession equations at
2PN order \cite{1995PhRvD..52..821K,2008PhRvD..78d4021R} for $q=1$,
\begin{align}
  \frac{d \mathbf{S_1}}{dt} = \frac{1}{2r^3}\left(7 \mathbf{L} -
  \frac{3}{2}M^2\xi\mathbf{\hat L} +\mathbf{S_2}\right) \times
  \mathbf{S_1} \;,
  \label{spinpre1}
  \\
  \frac{d \mathbf{S_2}}{dt} = \frac{1}{2r^3}\left(7 \mathbf{L} -
  \frac{3}{2}M^2\xi\mathbf{\hat L} +\mathbf{S_1}\right) \times
  \mathbf{S_2}\;.
\label{spinpre2}
\end{align}
imply $dS^2/dt =  2\mathbf{S}\cdot d\mathbf{S}/dt
\propto \mathbf{S} \cdot d(\mathbf{S}_1+\mathbf{S}_2)/dt
\propto \mathbf{S}\cdot (\mathbf{L}\times \mathbf{S})=0$.
It follows that for all $q=1$
configurations, the magnitude $S$ is conserved on the precession
and the radiation-reaction timescales [though not necessarily on the
orbital timescale over which
Eqs.~(\ref{spinpre1}-\ref{spinpre2}) are averaged].
This point was realized at least
as early as 2008 in Ref.~\cite{2008PhRvD..78d4021R}. A similar conclusion had previously been reached in \cite{1994PhRvD..49.6274A} with an incomplete set of 2PN equations and using some further assumptions. 

By cosine rule, the angle $\theta_L$ between orbital and total
angular momentum satisfies $2JL\cos\theta_L=J^2+L^2-S^2$,
so that with Eq.~(\ref{defxiq1}) and
$\mathbf{J}\cdot\mathbf{L}=L^2+\mathbf{S}\cdot
\mathbf{L}$ we find
\begin{align}
  J= \sqrt{L^2+S^2+L\xi M^2}\,.
  \label{Jexpr}
\end{align}
This relation holds on $t_{\rm pre}$ and, more importantly, on $t_{\rm RR}$
and therefore describes the evolution of $J$
as the separation decreases under GW emission.

These results can also be found using the precession-averaged
approach of \cite{2015PhRvL.114h1103K,2015PhRvD..92f4016G}:
With $q=1$, Eq.~(13) of \cite{2015PhRvD..92f4016G} gives
\begin{equation}
  \xi=\frac{J^2-L^2-S^2}{M^2L}\,,
  \label{eq:xiofSq1}
\end{equation}
and the effective potentials of BH spin precession, $\xi_+$ and $\xi_-$,
coincide for $q=1$, implying that $S$ is constant on the precession
time.
The precession averaging in
Eq.~(38) of \cite{2015PhRvD..92f4016G} then becomes a trivial operation,
so that
\begin{equation}
  \frac{dJ}{dL}=\frac{J^2+L^2-S^2}{2LJ}=\frac{2L+\xi M^2}{2J}
        ~~~~~\Rightarrow~~~~~2J\,dJ=2L\,dL+\xi M^2\,dL\,,
\end{equation}
and after integration $J^2-L^2-\xi LM^2=\mathrm{const}$ on $t_{\rm RR}$.
By Eq.~(\ref{eq:xiofSq1}), this constant must be $S^2$ and we have recovered
Eq.~(\ref{Jexpr}).

Note the remarkable character of this finding. With $q=1$ and
chosen parameters $S_1$, $S_2$, BH binaries are specified by pairs
$(\xi,S)$: both these quantities are constant on the precession and radiation reaction time
scale and uniquely determine the binary's characteristics.
$L$ is merely a measure for the binary separation $r$ and
$J$, the only evolving dependent variable, is
determined by the simple analytic expression (\ref{Jexpr}).
All other properties of the binary, such as the mutual orientation
of the BH spins and that of the orbital plane, follow from
straightforward geometric considerations of the triangles
$(\mathbf{J},\mathbf{L},\mathbf{S})$ and
$(\mathbf{S},\mathbf{S}_1,\mathbf{S}_2)$ as illustrated below. 
In contrast, for $q\ne 1$, PN integrations need to be initialized  with either $\xi$, $S$ \emph{and} $J$ at finite separation, or through
$\kappa_{\infty} = \lim_{r\to \infty}
\mathbf{S}\cdot\mathbf{\hat L}$ at infinitely large separation. 
One then needs to precession average $S^2$ using $dS/dt$ from Eq.~(26) of
\cite{2015PhRvD..92f4016G} and numerically integrate
for the evolution of $J$ on $t_{\rm RR}$ according to
Eq.~(38) of that work.

\subsection{Orbital-plane and spin precession}
\label{orbitalplaneandspinprec}

For $q\ne 1$ the precession dynamics are
conveniently parametrized by $S$, but we have seen that
in the equal-mass case
$S$ is constant and, hence, no longer suitable for this purpose.
Instead, we consider $\varphi'$ defined as the angle traced out
relative to some reference value $\varphi'_0$ by the spin
$\mathbf{S}_1$ in the plane orthogonal to the total spin
$\mathbf{S}$; see~Fig.~\ref{angles}.
We can fix the reference
value $\varphi_0'$ by defining  \cite{2015PhRvD..92f4016G}
\begin{align}
  \cos\varphi'
  = \frac{\mathbf{\hat{S}}_1\cdot \mathbf{\hat{S}}_{\bot}}
        {|\mathbf{\hat{S}}_1\times\mathbf{\hat{S}}|}
  = \frac{\mathbf{\hat S_1} \cdot \left[\left(\mathbf{\hat
  S} \times \mathbf{\hat L} \right) \times \mathbf{\hat
  S}\right]}{|\mathbf{\hat S_1}\times\mathbf{\hat S}|\,
        |\mathbf{\hat{S}}\times \mathbf{\hat{L}}|}\,,
        \label{eq:cosvarphi}
\end{align}
where $\mathbf{\hat{S}}_{\bot}$ is the unit vector perpendicular
to $\mathbf{S}$ and $\mathbf{S}\times\mathbf{L}$.
Note that the orientation of $\mathbf{S}_2$ in the same plane
is automatically determined through this definition by closure
of the triangle $(\mathbf{S},\,\mathbf{S}_1,\,\mathbf{S}_2)$.
The angle $\varphi'$ thus corresponds to rotations of  $\mathbf{S_1}$ and $\mathbf{S_2}$ about $\mathbf{S}$.
Using Eqs.~(7), (10), (28) and (29) of Ref.~\cite{2015PhRvD..92f4016G}
together with the spin-precession equations
(\ref{spinpre1}-\ref{spinpre2}) one can show that Eq.~(\ref{eq:cosvarphi})
implies
\begin{align}
  &\frac{d\,\cos\varphi'}{dt}
        =\frac{1}{|\mathbf{S}_1\times\mathbf{\hat{S}}|}
        \left( \mathbf{\hat{S}}_{\bot}\frac{d\mathbf{S}_1}{dt}
        + \mathbf{S}_1\frac{d\mathbf{\hat{S}}_{\bot}}{dt}\right)
\;\,  \Rightarrow \;\,\frac{d \varphi'}{dt} =-\frac{3 S}{r^3} \left(1- \xi
  \sqrt{\frac{M}{r}}\right) \leq 0\,,
  \label{eq:dvarphidt}
\end{align}
where the last inequality is manifest for separations $r\ge M$,
since $|\xi|<2S/M^2\le 1$ for Kerr BHs with $\chi\le1$. We conclude
that the angle
$\varphi'$ always evolves monotonically and $\cos \varphi'$ evolves periodically
back and forth between $-1$ and $+1$.
While the two spins precess about $\mathbf{S}$
with phase $\varphi'$, the orbital plane precesses about $\mathbf{J}$
with frequency \cite{2015PhRvL.114h1103K,2015PhRvD..92f4016G}
\begin{align}
  \Omega_z = \frac{1}{4}\frac{M^{3/2}}{r^{5/2}}\sqrt{1+\frac{\xi
  M^2}{L}+\frac{S^2}{L^2}} \left( 7 - \frac{3}{2} \frac{M^2 \xi
  }{L}\right)\,.
  \label{eq:Omegaz}
\end{align}
It is interesting to note that the timescale of these two phenomena
scale differently with the separation $r$. While the orbital plane
precesses on $t \sim \Omega_z^{-1}\propto r^{5/2}$, the two spins
precess about $\mathbf{S}$ on the longer time scale $t\sim \varphi'
dt/d\varphi'\propto r^{3}$. This appears surprising at first glance
since both
powers enter the orbit-averaged spin-precession equations
(\ref{spinpre1}-\ref{spinpre2}) and one would expect the shorter
timescale to dominate both features. For
generic $q\neq 1$ binaries, this is indeed the case;
both, $\Omega_z^{-1}$ and $ \varphi'
dt/d\varphi'$, scale as $r^{5/2}$ \cite{2015PhRvD..92f4016G}.
The markedly different behaviour of $q=1$ binaries arises
from a cancelation of all terms $\propto L \propto \sqrt{r}$
in the numerator on the right-hand side of Eq.~(\ref{eq:dvarphidt}).
The leading order term $\propto r^{-5/2}$ thus drops out of
the evolution of the two spins about $\mathbf{S}$,
but remains present in the precessional motion of $\mathbf{S}$ and
$\mathbf{L}$ about $\mathbf{J}$ as described by $\Omega_z$ of
Eq.~(\ref{eq:Omegaz}) where no such cancelation occurs.

\subsection{Spin morphologies}

The angle $\varphi'$ is a valuable quantity to mathematically formulate
the precession dynamics, but is not ideal for forming an intuitive
picture. This is achieved more conveniently
using instead the angles between the
vectors $\mathbf{S_1}$, $\mathbf{S_2}$ and $\mathbf{L}$,
\begin{align}
  \cos\theta_1 &= \mathbf{\hat S_1} \cdot \mathbf{\hat L}\,,
  \label{eq:cth1} \\
  \cos\theta_2 &= \mathbf{\hat S_2} \cdot \mathbf{\hat L}\,,
  \\
  \cos\theta_{12} &= \mathbf{\hat S_1} \cdot \mathbf{\hat S_2}\,,
  \label{eq:cth12}
\end{align}
and the azimuthal angle between the
projections of the two spins onto the orbital plane
\begin{align}
  \cos\Delta\Phi = \frac{ \mathbf{\hat S_1} \times \mathbf{\hat
  L}}{|\mathbf{\hat S_1} \times \mathbf{\hat L}|} \cdot \frac{
  \mathbf{\hat S_2} \times \mathbf{\hat L}}{|\mathbf{\hat S_2} \times
  \mathbf{\hat L}|}\,.
  \label{deltaphidef}
\end{align}
In general,
all four angles, $\theta_1$, $\theta_2$,  $\theta_{12}$ and $\Delta\Phi$,
oscillate on the precessional timescale, while
GW emission drives the
secular evolution. In the unequal-mass case, the angles
$\theta_1$, $\theta_2$ and $\theta_{12}$ evolve monotonically during
each precession cycle
(we define a cycle in this context to
cover the evolution of $\varphi'$ over an interval $\Delta \varphi'=\pi$),
while the evolution of $\Delta\Phi$ follows either of
three qualitatively different scenarios (see e.g. Fig.~3 of
\cite{2015PhRvD..92f4016G}):
\begin{enumerate}
  \item $\Delta\Phi$ can circulate spanning the full allowed range $[-\pi,\pi]$,
  \item $\Delta\Phi$ can librate about $0$, and never reaches $\pm \pi$,
  \item $\Delta\Phi$ can librate about $\pm \pi$, and never reaches $0$.
\end{enumerate}
This behavior
enables us to classify the precessional dynamics into \emph{morphologies}.
In general, the specific morphology of a binary is a function of
the separation $r$: radiation reaction may cause a BH
binary to transition from one morphology to another. These transitions
can only happen if either $\cos\theta_1=\pm 1$ or $\cos\theta_2
=\pm 1$ at some point during a precession cycle. At that moment,
one of the spins is (anti-) aligned with $\mathbf{L}$ and $\Delta\Phi$
is not well defined; cf.~Eq.~(\ref{deltaphidef}).

The $q=1$ case addressed here differs from this picture in some
important aspects. 
The angles describing the directions of the spins are obtained
from setting $q=1$ in Eqs.~(10) of \cite{2015PhRvD..92f4016G}
and evaluating the corresponding scalar products which gives
\begin{align}
  \cos\theta_1&= \frac{1}{4 S_1 S^2} \bigg[\xi M^2 (S^2+S_1^2-S_2^2)\notag\\
  &+ \sqrt{4S^2 - \xi^2
  M^4}\sqrt{S^2-(S_1-S_2)^2}\sqrt{(S_1+S_2)^2-S^2}\cos\varphi'\bigg]\,,
  \label{eq:cth1gen}
  \\
  \cos\theta_2&= \frac{1}{4 S_2 S^2} \bigg[\xi M^2 (S^2+S_2^2-S_1^2) \notag\\
  &- \sqrt{4S^2 - \xi^2
  M^4}\sqrt{S^2-(S_1-S_2)^2}\sqrt{(S_1+S_2)^2-S^2}\cos\varphi'\bigg]\,,\\
  \cos\theta_{12}&= \frac{S^2- S_1^2 - S_2^2}{2 S_1 S_2}\,,
  \label{eq:cth2gen}
  \\
  \cos\Delta\Phi&= \frac{\cos\theta_{12} - \cos\theta_1
  \cos\theta_2}{\sin\theta_1\sin\theta_2}\,.\label{eq:Dphigen}
\end{align}
We see that $\theta_{12}$ is constant
while the angles $\theta_1$ and $\theta_2$ oscillate between the two extrema
\begin{align}
  \cos\theta_{1\pm}&= \frac{1}{4 S_1 S^2} \bigg[\xi M^2
  (S^2+S_1^2-S_2^2)\notag\\
  &\pm \sqrt{4S^2 - \xi^2
  M^4}\sqrt{S^2-(S_1-S_2)^2}\sqrt{(S_1+S_2)^2-S^2}\bigg]\,,
 \label{ct1pm} \\
  \cos\theta_{2\pm}&= \frac{1}{4 S_2 S^2} \bigg[\xi M^2 (S^2+S_2^2-S_1^2)
  \notag\\
  &\mp \sqrt{4S^2 - \xi^2
  M^4}\sqrt{S^2-(S_1-S_2)^2}\sqrt{(S_1+S_2)^2-S^2}\bigg]\,,
  \label{ct2pm}
\end{align}
as $\varphi'$ evolves monotonically between $0$ and $\pi$ during
each precession cycle.
For $q= 1$,
the boundaries $\theta_{1\pm}$ and $\theta_{2\pm}$ do not depend
on either $J$ or $L$, but only on
the constants of motion $\xi,S, S_1$ and $S_2$. In contrast to the
$q\ne 1$ case, they therefore do not vary on
the radiation reaction timescale. This implies that the spin
morphology, i.e.~the qualitative evolution of $\Delta\Phi$, remains
unchanged at all separations. Equal-mass binaries do {\it not} exhibit
morphological transitions, which sets them qualitatively apart from
their generic unequal-mass counterparts.

The three different morphologies are still present, however, in the
$(\xi,\,S;\,S_1,\,S_2)$
parameter space of equal-mass binaries.  In particular, binaries in the two librating morphologies exist even at infinitely large separations ($r\to \infty$), while for $q\neq 1$ all binaries circulate in this limit. The main point is that a given binary never crosses the boundary between the different morphologies. 
These boundaries are given by the condition
$\cos\theta_{i\pm}=\pm1$ and a binary that happens to be sitting
at such a point will sweep through an
aligned configuration $\mathbf{S_i}\parallel \mathbf{L}$ during
each and every precession cycle throughout its entire inspiral.

The condition $\cos \theta_1 = \pm 1$ can be solved for $\xi$ as a function of $S$, yielding
\begin{align}
\xi=\pm \frac{2S}{M^2} \frac{ 2 S_1 S (S^2 +S_1^2 - S_2 ^2)
        + 
[4S_1^2 S_2^2 - (S^2 -S1^2 -S_2^2)^2]       
         |\cos\varphi'|
        \sqrt{\cos^2\varphi' - 1}}{4 S^2 S_1^2+
[4S_1^2 S_2^2 - (S^2 -S1^2 -S_2^2)^2]       
        (\cos^2\varphi'-1)},
\end{align}
while  $\cos \theta_2 = \pm 1$  has solutions
\begin{gather}
  \xi=\pm \frac{2S}{M^2} \frac{ 2 S_2 S (S^2 -S_1^2 + S_2 ^2)
        +
[4S_1^2 S_2^2 - (S^2 -S1^2 -S_2^2)^2]             |\cos\varphi'|
        \sqrt{\cos^2\varphi' - 1}}{4 S^2 S_2^2+
[4S_1^2 S_2^2 - (S^2 -S1^2 -S_2^2)^2]     
(\cos^2\varphi'-1)}.
\end{gather}
Real valued solutions only exist for the three discrete
values $\cos\varphi'=0,\,-1,\,+1$. For $\cos\varphi'=0$, we obtain
\begin{align}
\cos\theta_1=\pm1\quad\Rightarrow\quad  \xi &= \pm \frac{4S_1S^2}{M^2(S^2+S_1^2-S_2^2)}\,,
  \\ 
\cos\theta_2=\pm1\quad\Rightarrow\quad  \xi &= \pm \frac{4S_2S^2}{M^2(S^2-S_1^2+S_2^2)}\,.
\end{align}
It is straightforward to verify that these solutions violate the bound (\ref{defxiq1}) and can be discarded as unphysical.  On the other hand, the solutions for $|\cos\varphi'|=1$ 
[corresponding to $\theta_{i\pm}$ of Eq.~(\ref{ct1pm}-\ref{ct2pm})]
\begin{align}
\cos\theta_1=\pm1\quad\Rightarrow\quad  \xi &= \pm \frac{S^2 + S_1^2 - S_2^2}{ S_1}\,,
  \label{eq:xisol1}\\ 
\cos\theta_2=\pm1\quad\Rightarrow\quad  \xi &= \pm \frac{S^2 - S_1^2 + S_2^2}{ S_2}\,,
  \label{eq:xisol2}
\end{align}
fall into the allowed range (\ref{defxiq1}). These configurations correspond to binaries for which the spin
morphology is ill-defined during the entire inspiral and they
mark the boundary between the different morphologies as we will
discuss in more detail in the next section.

\section{Results: a simple picture}
\label{results}

\begin{figure}\centering
  \includegraphics[width=0.67\textwidth]{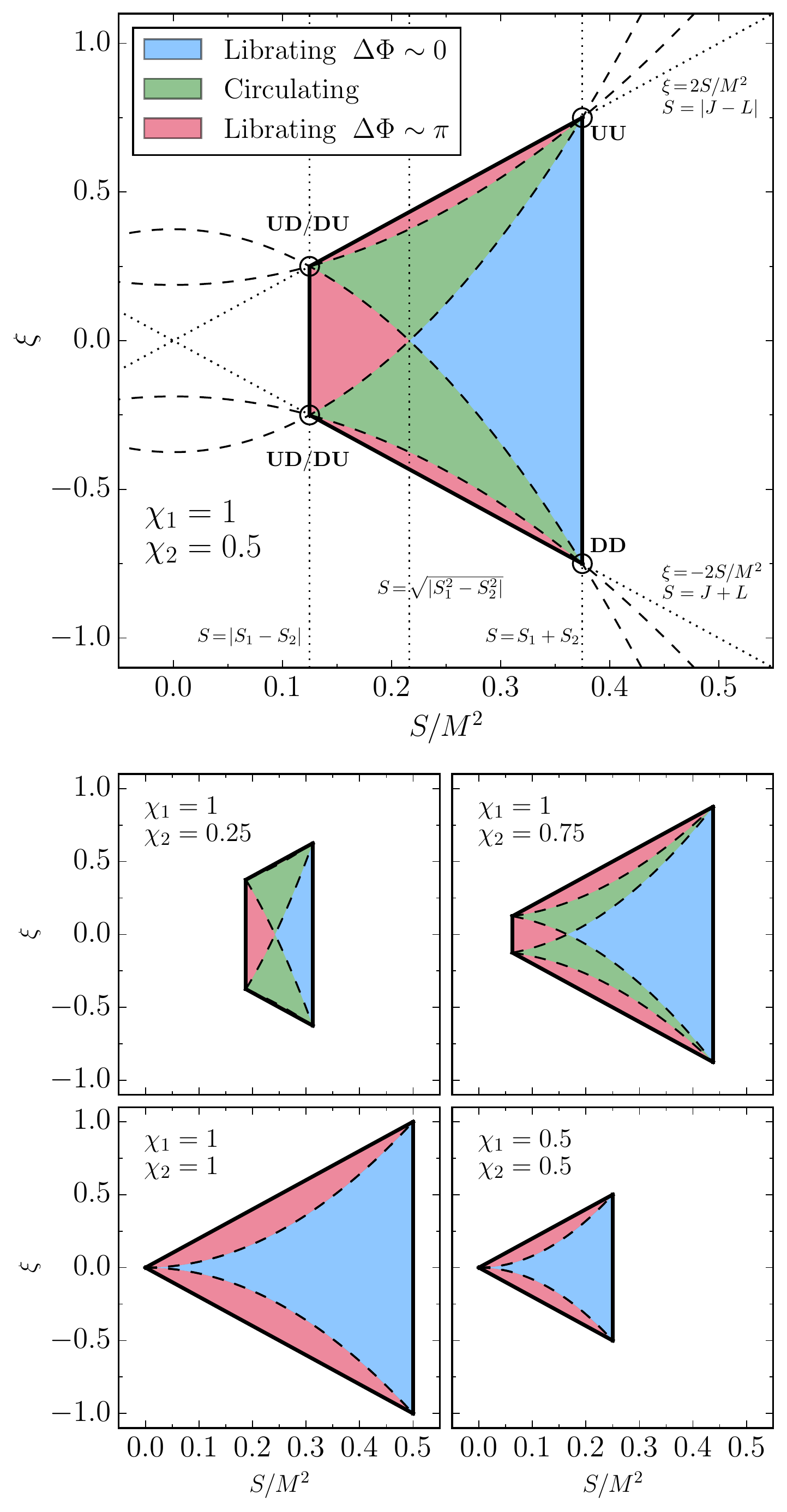}
  \caption{Configuration space of equal-mass BH binaries in the PN regime.
  Each binary is characterized by the two constants of motions $S$ and
  $\xi$, reported on the $x$ and $y$ axes respectively. The various panels
  show the configuration space for different values of the dimensionless
  spin magnitudes $\chi_1$ and $\chi_2$. As expected, the panels are
  invariant under a relabelling of the binary constituents
  $(\chi_{1},\chi_{2})\to(\chi_{2},\chi_{1})$.
  The physically
  allowed region is given by the area inside
  the four lines $S=|S_1+S_2|$
  and $\xi=\pm 2S/M^2$, shown as dotted curves in the top panel.
  In each panel, we show the resulting boundaries
  as a solid thick line and they
  mark the spin-orbit resonances, while round circles at the
  four corners {\bf UU}, {\bf DD} and {\bf UD/DU} mark binaries with
  both spins parallel to the orbital angular momentum. The spin morphology
  is encoded by the color of the shaded region and their boundaries
  are given by the dashed curves given by Eqs.~(\ref{eq:xisol1}-\ref{eq:xisol2}).
  }
  \label{xiSplane}
\end{figure}

In the previous section we have seen that the evolution of
a quasi-circular 
equal-mass binary
with fixed spin magnitudes $\chi_1$, $\chi_2$ is completely determined
by the values of $\xi$ and $S$ which remain constant during the inspiral.
The orbital angular momentum measures the binary separation and
the total angular momentum $J$ is given by Eq.~(\ref{Jexpr}). We can
therefore graphically visualize the set of all binaries with given
$\chi_1$, $\chi_2$ in the $(S,\xi)$  configuration space 
 and analyse the properties
of a binary as a function of its location in this diagram.
The resulting diagrams are shown in Fig.~\ref{xiSplane} for
several representative choices of $\chi_1$ and $\chi_2$.
Since $S$ and $\xi$ are both constants of motion, BH binaries are
stationary in these plots as they inspirals towards merger. Each
panel in  Fig.~\ref{xiSplane} therefore encompasses all equal-mass
BH binary evolutions with given spin magnitudes $\chi_1$ and $\chi_2$.

The physically allowed region in the parameter space is determined
by the constraint
$\mathbf{S}=\mathbf{S_1}+\mathbf{S_2}$ and the limits (\ref{xiconstraints})
for $\xi$ which gives us a total of four conditions
\begin{align}
  &|S_1-S_2|\leq S \leq S_1+S_2\,, \\
  &-2 S\leq \xi M^2 \leq 2S\,.
  \label{xiconstraints}
\end{align}
The resulting curves are shown as dotted lines in the top panel
of Fig.~\ref{xiSplane}.
Note that the condition
(\ref{xiconstraints}) for $\xi$ can be shown to be equivalent to the
constraint $\mathbf{S}=\mathbf{J}-\mathbf{L}$ for the magnitude
$S$; in particular $\xi = 2 S/M^2$ corresponds to $S=|J-L|$ and
$\xi = -2 S/M^2$ corresponds to $S=J+L$.

The four corners of the resulting allowed region in the ($S,\xi$) plane
correspond to binaries with both spins (anti-) aligned with the orbital
angular momentum (i.e.~$\sin\theta_1=\sin\theta_2=0$). More specifically,
the top-right (bottom-right) corner maximizes
(minimizes) $\xi$ and, hence, correspond to both spins being
aligned (antialigned) with
$\mathbf{L}$. We refer to these binaries as \emph{up-up}  ({\bf UU})
and $\emph{down-down}$ ({\bf DD}), respectively. The left boundary
of the allowed region minimizes $S$, so that the two spins $\mathbf{S}_1$
and $\mathbf{S}_2$ are antialigned with each other. The two left corners
represent the corresponding maximum (minimum) in $\xi$ where the
larger (smaller) spin is aligned and the other spin antialigned
with $\mathbf{L}$. We refer to these points as
\emph{up-down} or \emph{down-up} ({\bf
UD/DU}). Since $S$ is constant, all these four ``corner'' configurations are stable
under spin precession and phenomena like the \emph{up-down} instability
found in \cite{2015PhRvL.115n1102G} do not occur for equal-mass
binaries. Using Eq.~(2) in that paper, one immediately sees that \emph{both} instability thresholds $r_{\rm ud\pm}$ go to $\infty$ as $q\to 1$.

The angles of Eqs.~(\ref{eq:cth1})-(\ref{deltaphidef})
describing the mutual orientation of $\mathbf{S_1}$, $\mathbf{S_2}$
and $\mathbf{L}$ are all constant for binaries located on the edge
of the allowed region because all terms $\propto \cos \varphi'$ in
Eqs.~(\ref{eq:cth1gen}), (\ref{eq:cth2gen}) vanish for
either $S=|S_1 \pm S_2|$ or $\xi=\pm
2 S/M^2$. We note that
for $S=S_1+S_2$ the two spins are aligned with each other, so that
$\Delta \Phi=0$, while $\Delta \Phi=\pm \pi$ for the other cases $S=|S_1-S_2|$ and $\xi=\pm 2 S/M^2$. These configurations lying at the edge of the allowed region correspond to the spin-orbit resonances
discovered in Ref.~\cite{2004PhRvD..70l4020S}. The three momenta
$\mathbf{S_1}$, $\mathbf{S_2}$ and $\mathbf{L}$ share a common plane (i.e.~$\sin\Delta\Phi=0$)
 and jointly  precess about $\mathbf{J}$
with fixed mutual directions. While for $q\neq1$ such mutual directions
undergo secular changes due to radiation reaction, they are
truly constant for the $q=1$ case examined here (i.e.~they are 
independent of $L$). 
These are indeed the two families of resonant
solutions identified in \cite{2004PhRvD..70l4020S}, characterized by either $\Delta\Phi=0$  or $\Delta\Phi=\pm\pi$. 
The $\Delta\Phi=0$ family runs
from {\bf UU} to {\bf DD} along the right border where $S=S_1+S_2$, while the $\Delta\Phi=\pm \pi$ family  connects  {\bf UU} to {\bf DD} along the bottom ($\xi=-2S/M^2$), left  ($S=|S_1-S_2|$) and top ($\xi=2S/M^2$) borders. 

Next, we consider the different spin morphologies which we
display in Fig.~\ref{xiSplane} by color coding different areas
in the parameter space. 
Through Eqs.~(\ref{eq:cth1gen})-(\ref{eq:Dphigen}), we can regard $\Delta \Phi$ as a function of $\cos \varphi'$.
The three vectors $\mathbf{L}$, $\mathbf{S_1}$ and $\mathbf{S_2}$ are coplanar at $\cos \varphi'=\pm1$, and therefore  $\sin \Delta \Phi(\cos\varphi'\!\!=\!\!\pm1) = 0$.
A morphology boundary is defined by the discontinuous
change of the function $\Delta \Phi (\cos \varphi')$ as the parameters $S$, $\xi$ are varied a little. By
Eq.~(\ref{eq:Dphigen}), such a discontinuous change is only possible
at $\sin\theta_1 \times \sin\theta_2=0$ which are precisely
the solutions (\ref{eq:xisol1}-\ref{eq:xisol2}) shown as dashed parabolae in Fig.~\ref{xiSplane}.
We already know the behavior of the binaries on the edge of the
physically allowed region, so that
binaries located close to the edge boundaries
librate about either $\Delta\Phi=0$ (right; blue colored area
in Fig.~\ref{xiSplane}) or
$\Delta\Phi=\pm \pi$ (bottom, left, top; red colored) as they approach the
two families of resonant binaries where $\Delta\Phi={\rm constant}= 0,\pm \pi$. In the central regions (green),
binaries circulate in the full range $\Delta\Phi\in[-\pi,\pi]$.
Two
of the four solutions (\ref{eq:xisol1}), (\ref{eq:xisol2}) meet
at each of the four corners where both spins are (anti-) aligned with the
orbital angular momentum. These curves also
intersect each other at the special configuration
$S=\sqrt{|S_1^2-S_2^2|}$ and $\xi=0$, where two of the four \emph{instantaneously
aligned} configurations $\cos\theta_1=\pm 1$, $\cos\theta_2=\pm 1$ are touched  during each  precession
cycle.

The fractions of the parameter space belonging to each morphology
change with $\chi_1$ and $\chi_2$. In particular, more binaries
are allowed to circulate (librate) if the two spin magnitudes are
different (similar) to each other. In the limiting  case $\chi_1=\chi_2$
the four solutions of Eq.~(\ref{eq:xisol1}-\ref{eq:xisol2}) correspond
to only two distinct curves and binaries are not allowed to circulate.

It is trivial to show that the entire description we provided remains
unchanged under the inversion $(\chi_{1},\chi_{2})\to(\chi_{2},\chi_{1})$.
Finally, we point out that Fig.~\ref{xiSplane} in this paper should
\emph{not} be viewied as the $q=1$ equivalent of Fig.~4 in
\cite{2015PhRvD..92f4016G}: that figure merely represents
snapshots in the $(S,\xi)$ parameter space of a set of binaries with the same value of $J$ at a
given separation $r$.  
Our Fig.~\ref{xiSplane} instead displays all binaries
(with fixed $\chi_1$, $\chi_2$) over the entire PN  inspiral.

\section{Conclusions}
\label{discussion}

We have analyzed the dynamics of spinning equal-mass BH binaries
in light of the precession-averaged approach put forward in
\cite{2015PhRvL.114h1103K,2015PhRvD..92f4016G}. The existence of
an additional constant of motion, namely the magnitude of the total
spin $S=|\mathbf{S_1} + \mathbf{S_2}|$, greatly simplifies the PN
dynamics. For given spin magnitudes $S_1$ and $S_2$, PN inspirals
can be labelled by couples $(S,\xi)$, where $\xi$ is the projected
effective spin and is also a constant of motion. This entirely
determines the binary evolution at 2PN order of the spin precession
equations. 
The inspiral can be parameterized by
the magnitude of the orbital angular momentum
$L=m_1m_2\sqrt{r/M}$ and
the magnitude of the total angular
momentum $J=|\mathbf{L}+\mathbf{S}|$  is given by the analytic
expression of Eq.~(\ref{Jexpr}). The spin tilts oscillate between
the extrema given in Eqs.~(\ref{ct1pm}), (\ref{ct2pm}) which do not depend
on $L$ and, thus, on the binary separation.

Together, these features let us picture the entire parameter space
of equal-mass BH binaries using the diagrams of Fig.~\ref{xiSplane}.
While some features found for generic $q\neq 1$ binaries, such as
the existence the two families of spin-orbit resonances, persist
in the limit $q\to 1$, others turn out to be qualitatively different. In
particular, the spin morphology (i.e.~the qualitative evolution of
the spin orientation on the precessional timescale) is constant
throughout the inspiral and is uniquely determined by the values
$S$ and $\xi$ of the binary in question. As hinted in
\cite{2014PhRvD..89l4025G,2016PhRvD..93d4071T}, future high-significance
GW observations may provide direct measurements of the BH binary
spin morphology. In the case of (nearly) equal-mass events, this would
correspond to direct constraints on the spin directions at BH
formation; in contrast, for the $q\neq 1$ case
one needs to randomize over the
precessional phase and evolve the observed configurations back to
$r\to\infty$ \cite{2015PhRvD..92f4016G}.

Surprisingly, we found that precession of the BH spins and the
orbital plane takes place on different timescales if $q=1$: the
former is $\propto r^3$ and thus longer than the $\propto r^{5/2}$
result found for (i) the orbital plane precession in the $q=1$ case
and (ii) for both precession time scales for generic $q\neq 1$
binaries. In principle, this finding may allow for a further
timescale-averaging procedure, to
separate the evolution of the BH spins relative to the orbital
plane, and the evolution of the orbital plane in some inertial
reference frame (cf. \cite{ZhaoPrep}).

The results presented in this paper have been implemented in the
open-source python code {\sc precession}, available at
\href{http://www.davidegerosa.com/precession}{davidegerosa.com/precession}
\cite{2016PhRvD..93l4066G}. In particular, the constancy of $S$ and
Eq.~(\ref{Jexpr}) for the evolution of $J$ are   exploited explicitly
only if the code is run with $q=1$. For any value of $q<1$, the
general formalism of \cite{2015PhRvL.114h1103K,2015PhRvD..92f4016G}
is used. This  has been found to provide accurate results for $0.005
\lesssim q\lesssim 0.995$ \cite{2016PhRvD..93l4066G}.
As a possible future
extension of the code one may include the development of a hybrid approach
combining the two formulations.

\section*{Acknowledgments}
We thank Michael Kesden, Emanuele Berti and Richard O'Shaughnessy for several stimulating discussions.
DG is supported by NASA through Einstein Postdoctoral Fellowship
grant No. PF6-170152 awarded by the Chandra X-ray Center, which is
operated by the Smithsonian Astrophysical Observatory for NASA under
contract NAS8-03060. Additional support is acknowledged by NSF
CAREER grants PHY-1151197 and PHY-1404569, the UK STFC, and the Isaac Newton Studentship of the University of Cambridge. JV was supported by the Bridgewater Summer Undergraduate Research Opportunities Programme and the  Churchill College Small Grants fund at the University of Cambridge.
This work has received
funding from the European Union's Horizon 2020 research and innovation
programme under the Marie Sk\l odowska-Curie grant agreement No
690904, from H2020-ERC-2014-CoG Grant No.~"MaGRaTh" 646597,
from STFC Consolidator Grant No. ST/L000636/1,
the SDSC Comet, PSC-Bridges and TACC Stampede clusters through NSF-XSEDE Award
Nos.~PHY-090003,
the Cambridge High Performance
Computing Service Supercomputer Darwin using Strategic Research
Infrastructure Funding from the HEFCE and the STFC, and DiRAC's Cosmos
Shared Memory system through BIS Grant No.~ST/J005673/1 and STFC Grant
Nos.~ST/H008586/1, ST/K00333X/1.  Figures were generated
using the python package \textsc{matplotlib}
\cite{2007CSE.....9...90H}.

\newpage
\section*{References}

\bibliographystyle{iopart-num_davide}
\bibliography{equalmass}

\end{document}